# Two-photon-excited fluorescence (TPEF) and fluorescence lifetime imaging (FLIM) with sub-nanosecond pulses and a high analog bandwidth signal detection

Matthias Eibl[1], Sebastian Karpf[2], Hubertus Hakert[1], Daniel Weng[1], and Robert Huber[1]
[1]Institut für Biomedizinische Optik, Universität zu Lübeck, Lübeck, Peter-Monnik-Weg 4, 23562 Lübeck, Germany
[2]Department of Electrical Engineering, University of California, Los Angeles, California 90095, USA

## ABSTRACT

Two-photon excited fluorescence (TPEF) microscopy and fluorescence lifetime imaging (FLIM) are powerful imaging techniques in bio-molecular science. The need for elaborate light sources for TPEF and speed limitations for FLIM, however, hinder an even wider application. We present a way to overcome this limitations by combining a robust and inexpensive fiber laser for nonlinear excitation with a fast analog digitization method for rapid FLIM imaging.

The applied sub nanosecond pulsed laser source is synchronized to a high analog bandwidth signal detection for single shot TPEF- and single shot FLIM imaging. The actively modulated pulses at 1064nm from the fiber laser are adjustable from 50ps to 5ns with kW of peak power. At a typically applied pulse lengths and repetition rates, the duty cycle is comparable to typically used femtosecond pulses and thus the peak power is also comparable at same cw-power. Hence, both types of excitation should yield the same number of fluorescence photons per time on average when used for TPEF imaging. However, in the 100ps configuration, a thousand times more fluorescence photons are generated per pulse. In this paper, we now show that the higher number of fluorescence photons per pulse combined with a high analog bandwidth detection makes it possible to not only use a single pulse per pixel for TPEF imaging but also to resolve the exponential time decay for FLIM. To evaluate the performance of our system, we acquired FLIM images of a Convallaria sample with pixel rates of 1 MHz where the lifetime information is directly measured with a fast real time digitizer. With the presented results, we show that longer pulses in the many-10ps to nanosecond regime can be readily applied for TPEF imaging and enable new imaging modalities like single pulse FLIM.

## 1. INTRODUCTION

Two-photon excited fluorescence (TPEF) microscopy is a powerful technique in bio-molecular imaging to gain a greater understanding of processes ongoing on a cellular level[1]. The chemical composition of an investigated sample can be identified by discriminating different fluorophores and the nonlinear nature of the two-photon absorption effect allows for 3D deep tissue imaging [2]. Furthermore, even more information can be gained if the lifetime of fluorophores is recorded. Not only have the different fluorophores different lifetimes but more importantly, the lifetime changes with its neighboring molecules and thus can give information on the molecular bounds. With this wealth of information, TPEF microscopy combined with fluorescence lifetime imaging (FLIM) is a valuable tool for researchers in bio-molecular science [3-5].

However, a wider application of these imaging modalities is currently hindered by two drawbacks. First, TPEF microscopy is relying on elaborate excitation light sources. These light sources should provide high instantaneous power to induce two-photon absorption at low duty cycles to minimize the average power on the sample. The work horses for this kind of excitation are currently ultra-short pulse Ti:Sa lasers with pulse durations in the order of ~100 fs and a repetition rate of ~100 MHz [6]. As these lasers tend to be bulky, expensive, and not fiber compatible, TPEF imaging is usually done only in sophisticated optics labs. The second drawback when combining TPEF microscopy with FLIM is that the current gold standard for FLIM, time correlated single photon counting (TCSPC) [7], has certain speed limitations. Typical pixel rates of 10 kHz would lead to acquisition times of more than one and a half minutes for a 1024x1024 pixel image. For rapid assessment of large areas or for monitoring dynamic processes, this is often too slow.

In this paper, we present a way to overcome these two drawbacks. We combine a sub-nanosecond pulsed fiber laser with direct high analog bandwidth signal sampling. Fiber lasers are favorable over Ti:Sa lasers as they are very robust, alignment and maintenance free, and less expensive. Although they usually do not achieve such ultra-short pulses, they have been applied successfully in TPEF imaging [8-12] and it has been shown that the image quality compared to a Ti:Sa is the same if the same duty cycle is applied [13, 14]. The longer pulses also have the advantage that more fluorescence photons are produced per shot which makes it possible to derive the lifetime from a single excitation pulse already by recording the TPEF signal with a high analog bandwidth detection [13]. Furthermore, Giacomelli et al. showed that direct analog sampling of the fluorescence decay allows to rapidly assess large area biopsies from a surgical resection of malignant tissue [15]. With these two measures combined, the easy-to-use fiber laser and high analog bandwidth detection of the TPEF signal, we show FLIM images with pixel rates of 1 MHz.

## 2. METHODS

A simplified scheme of our microscope setup is depicted in figure 1. We use a homebuilt fiber laser as excitation source. Its output beam is coupled into a homebuilt microscope setup and the fluorescence light is detected with a high analog bandwidth detection.

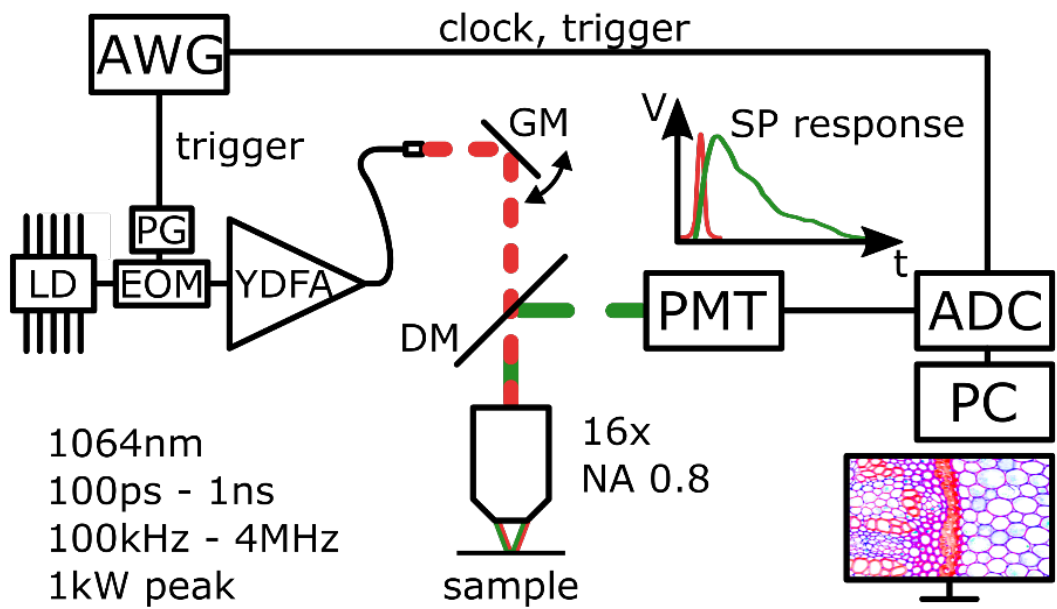

Fig. 1: Basic concept of the single pulse FLIM setup. An arbitrary waveform generator (AWG) provides trigger and timing signals for the laser pulse generation and the digitization. Excitation pulses are provided by an all fiber master oscillator power amplifier (MOPA) based on. Briefly, the output of a laser diode (LD) gets modulated to short pulses by an electro optical modulator (EOM), driven by a high speed pulse generator (PG). These typically 100 ps seed pulses with 1 MHz repetition rate are amplified by ytterbium doped fiber amplifiers (YDFA) to ~1 kW of peak power. Galvanometric mirrors (GM) are used to scan the high energy pulses over the sample. There, one single pulse has enough power and duration to excite many fluorophores so that many fluorescence photons can be detected by a photomultiplier tube (PMT). A fast time response of the PMT and high bandwidth of the analog to digital converter (ADC) ensure a high time resolution for accurate lifetime determination on a personal computer (PC).

The all fiber excitation laser is based on the master oscillator power amplifier (MOPA) concept. Briefly, the spectrally narrowband and continuous wave output of a laser diode (LD) gets amplitude modulated by an electro optical modulator (EOM, Photline, NIR-MX-LN 10). The electrical pulses to drive the EOM are provided by a high speed pulse generator (PG, Alnair, EPG-210) and can be adjusted continuously from 50 ps to 250 ps. After the EOM, the pulses with about 100 mW peak power are inserted in a multistage ytterbium doped fiber amplifier (YDFA) where they are amplified to peak powers of 1 kW. The peak output power is limited by the onset of stimulated Raman scattering (SRS) in the fiber. While this is usually an unwanted effect, it also can be harnessed to create other excitation wavelength as well [16-18].

These excitation pulses are coupled into a homebuilt multi-photon microscope which follows partly the design concept presented in [19]. The output of the single mode fiber is collimated (Thorlabs, C280TME-1064) to a beam diameter of 4.1 mm. Two galvanometric mirrors (GM, Thorlabs, GVS002) deflect the beam. The scanning plane is relayed to the back focal plane of the microscope objective with a 75 mm achromatic lens (Thorlabs, AC508-075-C-ML) as scan lens and a combination of a 300 mm and a 400 mm lens (Thorlabs, AC508-300-C-ML, TL2:400mm AC508-400-CML) as tubus lens. With this lens combination, the beam is magnified to a diameter of 12.3 mm. A NA 0.8 16x objective (Nikon, CFI75 LWD 16xW) is used to focus the beam onto the sample and collect the fluorescence light. Excitation and fluorescence light are separated by a dichroic mirror (Thorlabs, DMLP950R) before the objective and the fluorescence light is detected by a fast multi-alkali photomultiplier tube (PMT, Hamamatsu, H12056-20). To further suppress remaining excitation light, additional spectral filters are placed before the PMT.

The output of the fast PMT is directly connected to a high analog bandwidth high resolution digitizer (ADC, Alazartech, ATS9373). The 12 bit card has a bandwidth of 2 GHz, samples at a rate of 4 GS/s and streams the data directly into the PC's memory. The data acquisition is synchronized to the excitation laser by an inter-channel locked arbitrary waveform generator (AWG, TTi, TGA12104). This AWG provides trigger signals for the laser and trigger signals as well as a 10 MHz sample clock reference for the ADC. The clock reference gets multiplied on the ADC's onboard PLL (phase locked loop) to 2 GHz were the rising and falling edge are used to trigger data acquisition resulting in a 4GS/s sampling rate.

For each excitation pulse, 40 to 100 samples -an equivalent of 10 to 25 ns- are stored for further processing. For TPEF images, only the maximum intensity, i.e. the first few samples, are used for coloring. To form a FLIM image, the decay curve gets fitted to a lifetime model. This model function is a convolution of the measured instrument response function (IRF) and an exponential decay function. This approach provides a more robust lifetime fit result than the straight-forward way to de-convolute the measured lifetime data with the IRF and fit the resulting exponential decay. The resulting lifetime is mapped to a color coding using the HSV (hue, saturation, value) color space. Hue represents the lifetime and value the fluorescence amplitude. Saturation is usually not used but can be implemented to add a quality of fit parameter.

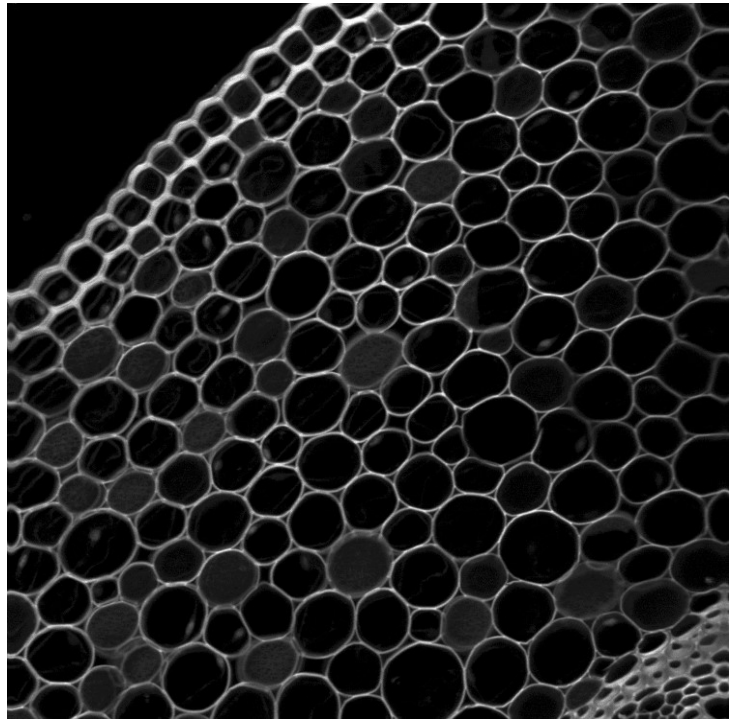

Fig. 2: TEPF image of an acridine orange stained *convalaria majalis* stem. Image size is about 0.5 mm x 0.5 mm at a resolution of 2048x2048 pixel. Acquired with 200 kHz pulse rate and a single pulse per pixel.

## 3. RESULTS

Figure 2 shows a TPEF image of a slice of *convalaria majalis* stem stained with acridine orange (Lieder GmbH). It has a size of 2048x2048 pixels and covers a region of ~0.5 mm x 0.5 mm. A single pulse excitation at a rate of 200 kHz with a pulse length of 500 ps was used. As reported previously, the image quality is comparable to commercially available two-photon microscopes [13].

For all following results, the laser repetition rate was set to 1 MHz and the pulse length to 100 ps. To use this setup for FLIM imaging, the time response of the system has to be characterized. An image of urea crystals was acquired and the SHG signal thereof analyzed. Figure 3 b) shows the TPEF intensity image of the crystals. It has a size of 512x512 pixels and was acquired within 670 ms. However, the actual raw data acquisition time was only 260 ms. The difference between the raw data acquisition time and the total acquisition time is due to the fact that we have not implemented bidirectional scanning and we use only 80% of our scan range to avoid nonlinear movement artefacts at the image edges. For each pixel in the image, 60 samples -which represents 15 ns- were acquired for further processing. The single pulse response of a representative pixel is shown in figure 3 a). This shows the fastest time response that can be acquired with our setup. A Gaussian fit of the measured data at has a FWHM of 1.34 ns. It should be noted, however, that this does not pose an ultimate limit to the shortest lifetime that can be identified with this technique. As the measured curve is a convolution of the actual lifetime with the IRF, one can extract shorter lifetimes than the above mentioned 1.34 ns. The limitation is rather given by the variation (standard deviation) of the width of the FWHM which was determined to be 63 ps over the whole image. If this limit can be reached has to be evaluated in further experiments, though.

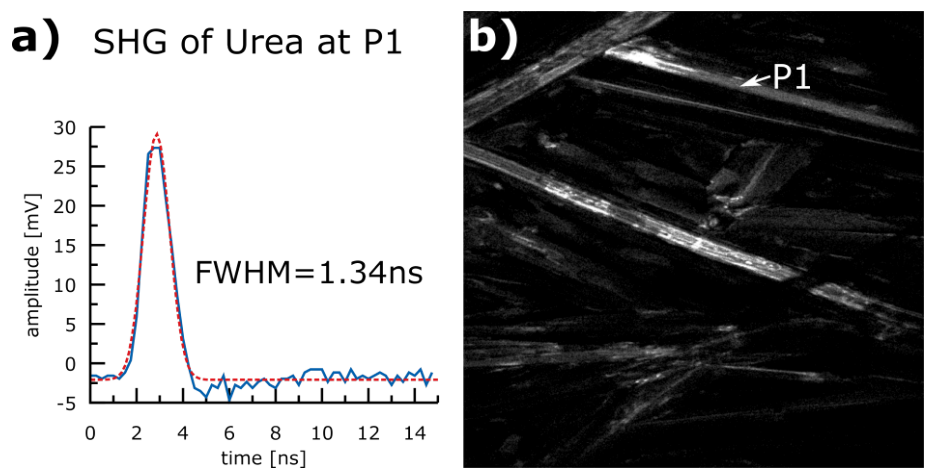

Fig. 3: Characterization of the instrument response function (IRF) with second harmonic signal (SHG) of urea crystals. **a)** SHG signal at pixel P1. Blue curve shows measured data, red curve shows a Gaussian fit. The FWHM of the fit is 1.34 ns. **b)** shows the TPEF image of the acquired Urea crystals. The image has a size of 512x512 pixels and covers a region of 0.5 mm x 0.5 mm. It was acquired with 1 MHz pixel rate and raw data acquisition time was only 260 ms.

Figure 4 shows FLIM images of a *convalaria majalis* stem stained with acridine orange (Lieder GmbH). The left image (a) ) was acquired with a single pulse per pixel excitation, resulting in a pixel rate of 1 MHz. With an image size of 512x512 pixels, this results in a raw data acquisition time of 260 ms. As mentioned above, the total acquisition time was 670 ms due to scanning limitations. The right image (b) ) covers the same region of about 0.5 mm x 0.5 mm. It was 4 times averaged and, therefore, shows an even higher image quality. For both images, the TPEF intensity is used as saturation value and the lifetime is mapped onto a color as indicated in the image. The color bar ranges from 0.5 ns to 2.4 ns and clearly three different regions can be identified within the image. The main structure is colored blue which indicates lifetimes in the region of ~1 ns, vascular bundles with lifetimes about 2 ns appear red, and within the main structures, green areas appear which are probably chloroplasts with a shorter lifetime of ~0.5 ns. Overall, different regions within in the convalaria test sample can be clearly distinguished by their lifetime. It shows that the time resolution is sufficient to differentiate three species with small lifetime span of only about 2 ns.

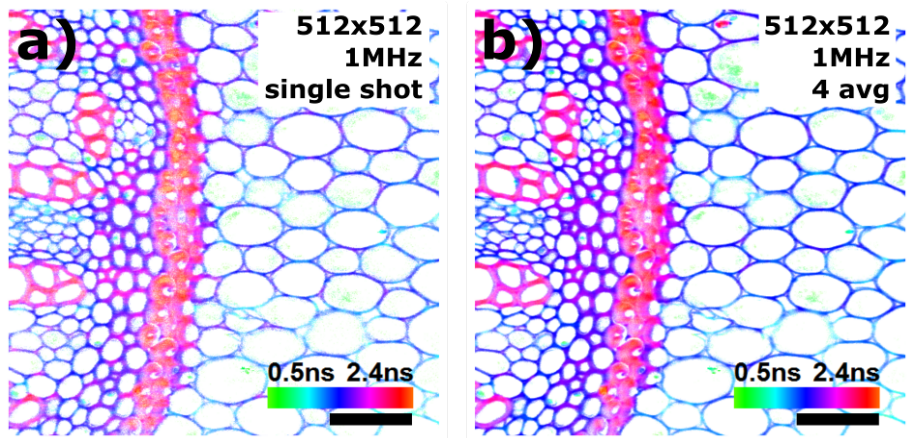

Fig. 4: FLIM images of a *convalaria majalis* stem stained with acridine orange (Lieder GmbH). Both images have 512x512 pixels and represent a region of ~0.5 mm x 0.5 mm. Scale bar is 100 µm. **a)** FLIM image with single pulse excitation per pixel. A pixel rate of 1 MHz was achieved with a data acquisition time of 260 ms. **b)** Same region four times averaged.

## 4. CONCLUSION AND OUTLOOK

The presented results show that longer pulses in the sub-nanosecond regime can be applied not only for TPEF imaging but also for fast FLIM imaging. The long and high intensity pulses induce a strong TPEF signal. The analog detection method makes it possible to process this high number of fluorescence photons to derive the fluorescence lifetime with a single excitation pulse already. We have shown a 1 MHz pixel rate and even higher pixel rates could be feasible. Despite the high speed, the lifetime determinate was accurate and robust which is vital if substantial information has to be extracted from the lifetime itself for example in Förster resonance energy transfer (FRET). The all fiber setup makes it possible to include the TPEF imaging and FLIM modality into a fiber endoscope. Furthermore, it can be combined with fast optical coherence tomography (OCT) [20-22]and even live display could be feasible with GPU lifetime calculation as was presented for OCT already [23, 24]. In total, this is a promising technology to bring TPEF and FLIM imaging to the workbench of researchers in bio-molecular science and to make live multi-modal endoscopic imaging with OCT possible.